\documentclass[runningheads]{llncs}

\usepackage[T1]{fontenc}
\usepackage{graphicx}
\usepackage{amsmath,amssymb}
\usepackage{algorithm}
\usepackage{algpseudocode}
\usepackage{multirow}
\usepackage{booktabs}
\usepackage{dsfont}  
\usepackage{algorithmicx}
\usepackage{CJK}
\usepackage{enumitem}
\setlist{leftmargin=0.2cm}
\newcommand \footnoteONLYtext[1]
{
	\let \mybackup \thefootnote
	\let \thefootnote \relax
	\footnotetext{#1}
	\let \thefootnote \mybackup
	\let \mybackup \imareallyundefinedcommand
}

\title{Towards Efficient Quantity Retrieval from Text: An Approach via Description Parsing and Weak Supervision}
\titlerunning{Efficient Quantity Retrieval via Parsing and Weak Supervision}

\author{Yixuan Cao\inst{1,2*} \and
Zhengrong Chen\inst{1,2} \and
Chengxuan Xia\inst{1,2} \and
Kun Wu\inst{1,2} \and
Ping Luo\inst{1,2,3}}

\authorrunning{Cao et al.}

\institute{Key Lab of Intelligent Information Processing of Chinese Academy of Sciences (CAS), Institute of Computing Technology, CAS, Beijing, China \\
\and University of Chinese Academy of Sciences, CAS, Beijing, China
\and Peng Cheng Laboratory, Shenzhen, China\\
\email{\{caoyixuan, luop\}@ict.ac.cn} \\ 
}
\begin{document}

\maketitle
\footnoteONLYtext{\textsuperscript{*}Corresponding author: Yixuan Cao.}
\footnoteONLYtext{The Version of Record of this contribution is published in DEXA 2025.}

\begin{abstract}
Quantitative facts are constantly produced in the operations of companies and governments, supporting data-driven decision-making. While common facts are structured, many long-tail quantitative facts remain hidden in unstructured documents, making them difficult to access. We propose the task of Quantity Retrieval—given a description of a quantitative fact, the system returns the relevant value and supporting evidence. Understanding quantity semantics in context is key to solving this task.
We introduce a framework based on description parsing, which converts text into structured (description, quantity) pairs and enables effective retrieval. To improve learning, we construct a large paraphrase dataset using weak supervision based on quantity co-occurrence. We demonstrate our approach using a large corpus of financial annual reports and a newly annotated quantity description dataset. Experiments show our framework significantly improves top-1 retrieval accuracy from 30.98\% to 64.66\%.

\end{abstract}

\section{Introduction}

\begin{figure}[t]
\centering
\includegraphics[width=.8\textwidth]{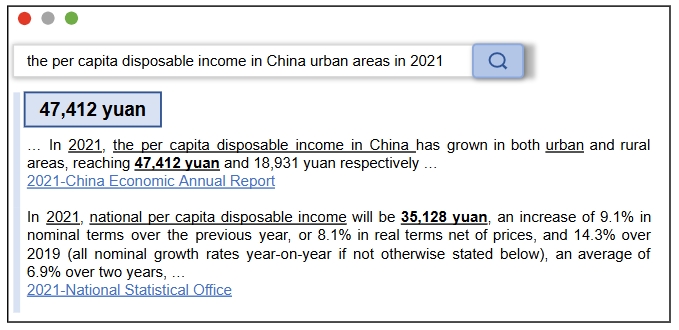}
\caption{A sketch of Quantity Retrieval Application.}
\label{fig:example}
\end{figure}
\vspace{-1em}

Quantitative facts are regularly produced by companies and governments and published in periodic reports. For example, companies disclose operational data such as sales, revenue, and employee count, while governments release economic indicators like GDP, market size, and investment figures. These quantities underpin decision-making in business, policy, and daily life.
To support such decisions, one must first locate the relevant quantities. The search query is typically a textual description, e.g., the number of cars delivered by Tesla in 2020, the GDP of the US in 2021, and the expected result is a value of 180,570 or \$22.99 trillion.

While popular quantities may be retrievable via commercial search engines, long-tail facts often are not. Many domain-specific quantities are buried in long, professional documents—often in PDF or scanned form—where sentence and paragraph structure is lost. Locating them requires domain knowledge and manual navigation through extensive documentation. For instance, financial analysis demands numerous related quantities, making the process laborious and time-consuming. Recent advances in document parsing~\cite{cao2021towards,li2020cross} have improved accessibility, enabling search engines to index such content more effectively.

In this work, we propose the task of \emph{Quantity Retrieval}: given a textual description of a quantitative fact, return relevant quantities and supporting evidence—a snippet from the document that describes the quantity. An example is shown in Figure~\ref{fig:example}, where the model returns a target quantity with contextual evidence and other relevant matches below. We outline the key challenges next.

% \textbf{Challenge 1: Fine-grained Information Need.} The information inside a document is myriad while the intended result is pinpoint. A professional document usually has hundreds of pages and contains abundant quantities and descriptions. For example, an annual report has over one hundred pages and five thousands quantities on average. On the contrary, quantitative search expects a pinpoint result, i.e. a quantity. Thus, except for the target quantities and descriptions, other information in the document will dilute the relatedness between the query and the document. This is different from other information retrieval tasks which retrieve at document level. 

\begin{figure*}[t]
\centering
\includegraphics[width=\textwidth]{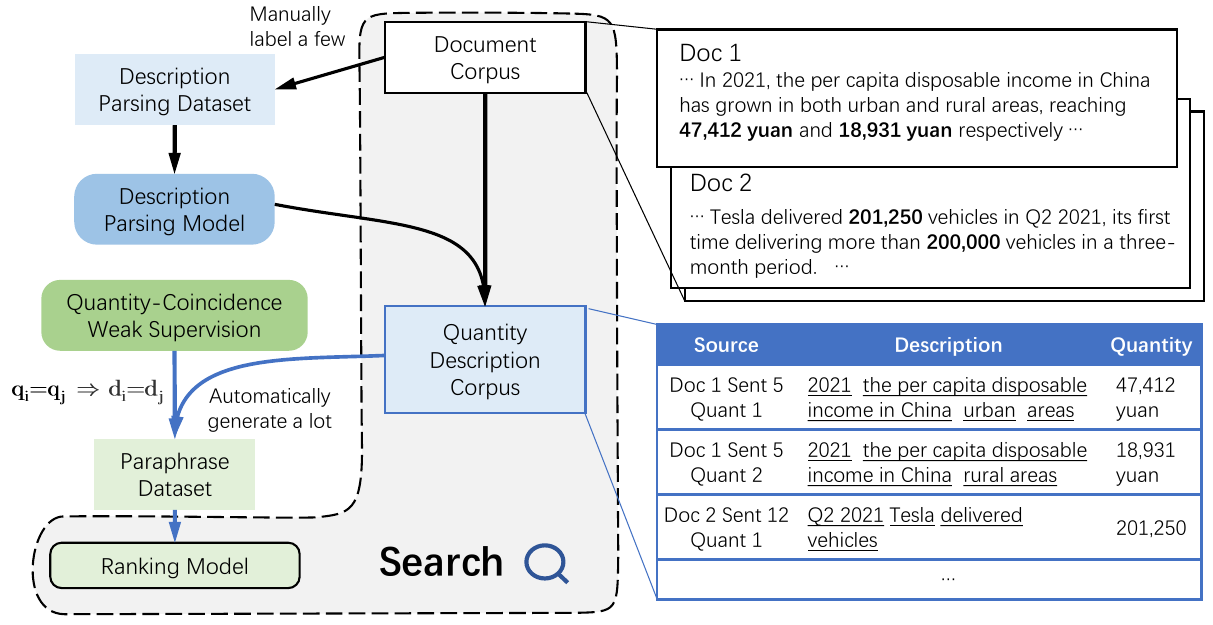}
\caption{The overview of our proposed quantity retrieval via description parsing (the blue parts) and weak supervision (the green parts).}
\label{fig:overview}
\end{figure*}

\textbf{Challenge 1: Textual vs. Semantic Similarity.}  
Textual similarity does not always reflect semantic correctness. Consider the query “\textit{number of R\&D staff of Apple in 2019}.” A sentence in Apple’s 2020 annual report—“In 2020, the number of R\&D staff in Apple was 10,000, increased by 25\% compared to 2019”—contains all the query terms but provides quantities for 2020 and the change from 2019, not the target value itself.

\textbf{Challenge 2: Paraphrases vs. Confusing Descriptions.}  
Different phrasings may refer to the same fact (e.g., “sales” vs. “market size” in “smartphone sales/market size in 2021”), while similar words may describe different facts (e.g., “R\&D expense” vs. “R\&D expenditure”). The latter differ in accounting context—expenses are tied to revenue, while expenditures imply long-term investment.

\textbf{Our Approach.}  
We address these challenges using NLP techniques. An overview is shown in Figure~\ref{fig:overview}. First, we convert raw documents into a \emph{quantity description corpus}, where each record pairs a description with a quantity. For example, in Doc1 of Figure~\ref{fig:overview}, two records are extracted. Each description is a concise textual span that explains what the quantity measures. We treat description parsing as a sequence labeling task and train a deep neural model.

Next, we construct a paraphrase dataset by grouping different descriptions that share the same quantity. A ranking model is trained on this weakly supervised data to measure description similarity. At retrieval time, given a query, the model finds the most similar description in the corpus and returns its quantity and supporting snippet.

To support this, we manually annotated 42,130 sentences and 112,117 quantities with descriptions. Our model achieves an 89.47 F1 score at the segment level—comparable to human performance. We further parsed a larger corpus, yielding around 3 million quantities. Experiments show that both our parsing method and weak supervision strategy significantly improve retrieval. These results highlight the value of deeper semantic analysis for fine-grained information retrieval.

In summary, the contributions of our work are as follows:
\begin{itemize}
    \item We propose the task of quantity retrieval from text.
    \item We propose a description parsing and weak supervision approach to understand the meaning of quantities in text.
    \item We collect a large-scale dataset, analyze the characteristics of the description of quantity, and conduct extensive experiments and analysis.
\end{itemize}

\section{The Description of a Quantity}\label{subsec:description}

% \begin{figure}[t]
% \centering
% \includegraphics[width=0.45\textwidth]{figs/parsing-examples.png}
% \caption{Examples of Description Parsing.}
% \label{fig:parsing_example}
% \end{figure}

This section defines quantity descriptions and highlights their key characteristics.

The meaning of a quantity in a sentence can be expressed through the text spans that describe it. If we rewrite a sentence to the form “$X$ is $Q$,” then $X$ represents the quantity’s meaning. We refer to $X$ as the \emph{description} of quantity $Q$. Examples appear in Figure~\ref{fig:overview} (right).

Quantity descriptions in text exhibit several properties. First, a sentence may contain multiple quantities. Second, a single description may consist of multiple, possibly discontinuous, text segments (see underlined examples in Figure~\ref{fig:overview}). Third, some quantities are not factual—for example, “200,000” in Doc2—which cannot be cast into the “$X$ is $Q$” form. These are excluded from description parsing and retrieval.

Descriptions are composed of several semantic factors:
\begin{itemize}
    \item \textbf{Performance indicator}: The core of the description—what the quantity measures (e.g., “delivered vehicles,” “per capita disposable income”).
    \item \textbf{Time}: Indicates when the measurement was taken; necessary for handling temporal queries.
    \item \textbf{Subject}: The entity the quantity is about (e.g., “Tesla”).
    \item \textbf{Place}: The geographic or regional scope (e.g., “China”).
\end{itemize}

Descriptions can be complex and include more factors. Lamm et al.~\cite{lamm2018qsrl}, for example, identify 11 roles for quantities in financial text. However, we focus only on the time factor, as distinguishing all possible factors is labor-intensive and not essential for our retrieval task.

We now formally define the description parsing task.

\begin{definition}
The \emph{Description Parsing} task is that given a sentence $s=(w_1, ..., w_n)$ and a set of quantities $\{q_i=(w_a, ..., w_b)\}_{i=1}^{m}$ in $s$, for each quantity $q$, output $d$, a continuous or dis-continuous sub-sequence of $s$ that fully and concisely describe $q$, in which ``fully'' requires all the necessary factors about $q$ are included and ``concisely'' means no words in the description could be removed.
\end{definition}

\section{Quantity Retrieval via Description Parsing and Weak Supervision}

\begin{definition}
The \emph{Quantity Retrieval} task is that, based on a document corpus $\mathcal{C}$, given a textual query $x$ which describes the required quantity, return a list of quantities and their evidence (text around quantities) in which quantities more related to $x$ appear earlier in the result list.
\end{definition}

Figure~\ref{fig:overview} illustrates the overall framework. The system begins with a document corpus $\mathcal{C}_d$. In the first stage (blue boxes), we train a description parsing model using a manually annotated dataset, and apply it to $\mathcal{C}_d$ to construct a quantity description corpus $\mathcal{C}_q = \{(d_i, q_i)\}_{i=1}^{n}$, where each record is a (description, quantity) pair.

Next (green boxes), we train a ranking model using weak supervision. We generate paraphrase pairs by identifying different descriptions that share the same quantity—based on the assumption that such descriptions are likely paraphrases. These pairs are used to train the ranking model for paraphrase detection.

At inference time, given a query, the ranking model retrieves the most similar descriptions from $\mathcal{C}_q$ and returns their corresponding quantities along with contextual snippets.

The remainder of this section details the description parsing, weak supervision strategy, and ranking model.

\subsection{Description Parsing Model}\label{subsec:parsing}

The description parsing model is fully supervised and requires an annotated dataset (described in Section~\ref{sec:exp-parse}). Here, we focus on the model design.

Since sentences may contain multiple, possibly overlapping quantities with discontinuous descriptions, parsing all of them simultaneously is challenging. We therefore adopt a one-at-a-time approach: the model parses the description of a single quantity per pass. This is similar to semantic role labeling~\cite{kamathsurvey}, but centered on quantities rather than predicates.

Our model follows a sequence labeling framework, with a modification to highlight the target quantity. Given an input sentence $s = (w_1, ..., w_n)$ and a pivot quantity $q = (w_i, ..., w_j)$, we insert special tokens to mark its span:  
$s' = (w_1, ..., \texttt{[START]}, w_i, ..., w_j, \texttt{[END]}, ..., w_n)$.  
A separate sentence is constructed for each quantity in a sentence.

The model has two layers: encoding and classification.

The \textbf{encoding layer} maps the input sequence to contextualized hidden vectors:
\begin{equation*}
    (h_1, h_{\texttt{[START]}}, h_i, ..., h_j, h_{\texttt{[END]}}, ..., h_n) = \text{Encoder}(s')
\end{equation*}

The \textbf{classification layer} assigns BIEO tags (Begin, Inside, End, Outside) to each token using a feed-forward network over the hidden vectors~\cite{yang2018design}, identifying segments that form the quantity description.

For example, in the second quantity from Doc1 in Figure~\ref{fig:example}, the description includes “2021,” “the per capita disposable income,” “China,” and “rural areas.” The tagged sequence is:

\begin{quote}
In$_O$ 2021$_B$ ,$_O$ the$_B$ per$_I$ capita$_I$ disposable$_I$ income$_E$ in$_O$ China$_B$ has$_O$ grown$_O$ in$_O$ both$_O$ urban$_O$ and$_O$ rural$_B$ areas$_E$ ,$_O$ reaching$_O$ 47,412$_O$ yuan$_O$ and$_O$ [START]$_O$ 18,931$_O$ yuan$_O$ [END]$_O$ respectively$_O$ .$_O$
\end{quote}

These tags are then used to reconstruct the quantity’s description.

\subsection{Value-Coincidence Weak Supervision}

\begin{algorithm}
\caption{Value-Coincidence Weak Supervision}
\label{algo:weak_supervision}
\begin{algorithmic}
\State Initialize BM25 using $\mathcal{C}_q$, return top $k$ results.
\State paraphrase = []
\State confusing = []
\For {$(d_i, q_i)$ in $\mathcal{C}_q$}:
    \For{ $(d_j, q_j)$ in BM25($d_i$, $k$)}:
        \If{ \texttt{SameValue}($q_i, q_j$)}
            \State paraphrase.append([$q_i, q_j$])
        \Else
            \State confusing.append([$q_i, q_j$])
        \EndIf
    \EndFor
\EndFor
            
\end{algorithmic}
\end{algorithm}

The \textit{paraphrase and confusing description challenge} discussed in the Introduction cannot be addressed by dependency parsing. An intuitive solution for this is to construct a paraphrase knowledge base.
But it heavily relies on professional expert knowledge. However, $\mathcal{C}_q$ provides us an opportunity to construct such a paraphrase dataset atomically.
This is based on two observations:
\begin{enumerate}
    \item Important paraphrases exist in the corpus. If a quantitative fact (or performance indicator) has multiple  descriptions, they will appear in the corpus. Otherwise, if there is only one description for a fact in a large corpus, users might tend to use this standard description during searching. This is tenable when the corpus is large enough.
    \item Paraphrases can be mined automatically. A characteristic of quantitative fact is that paraphrase descriptions share the same quantity value. So, we can mine paraphrases based on quantity-coincidence in $\mathcal{C}_q$.
Specifically, if two descriptions share the same quantity and are somewhat similar in text, they are likely to paraphrase referring to the same fact.
\end{enumerate}

We demonstrate the rationale for the second observation as follows.
For any two records $(d_i, q_i)$, $(d_j, q_j)$ referring to facts $f_i$ and $f_j$, we denote the event that $d_i$ is similar to $d_j$ as $D$, $q_i=q_j$ as $Q$, $f_i=f_j$ as $F$. The probability that these two records refer to the same fact, given the quantity is the same and the description is similar is defined as:
$$
\begin{aligned}
    &P(F|D, Q) = \frac{P(F, D, Q)}{P(D, Q)} = \frac{P(F)}{P(D, Q)} \\
    & = \frac{P(F)}{P(D, Q | F)P(F) + P(D, Q | \bar{F})P(\bar{F})}  \\
    &=  \frac{P(F)}{P(F) \left(P(D,Q|F)-P(D, Q | \bar{F})\right) + P(D, Q | \bar{F})} \\
    &\approx \frac{P(F)}{P(F) + P(D, Q | \bar{F})}   = \frac{1}{1+\frac{P(D,Q|\bar{F})}{P(F)}} =  \frac{1}{1+\frac{P(D|Q\bar{F})  P(Q|\bar{F})}{P(F)}} 
\end{aligned}
$$
Suppose there are N records in $\mathcal{C}_q$, and each fact has $r$ records with similar descriptions, then $P(F)=\frac{r}{N}$. The vocabulary size of the corpus is $V$ and the number of terms in a description is $l$, then  $P(D|Q,\bar{F}) \approx P(D|\bar{F}) \approx \frac{1}{V^l}$. A quantity has $s$ significant digits, then $P(Q|\bar{F}) = \frac{1}{10^s}$. Therefore, $\frac{P(D|Q\bar{F})P(Q|\bar{F})}{P(F)}=\frac{N}{V^l 10^s r} \rightarrow 0$ when V is large (like $10^4$) and N is not too large (like $10^6$), and thus $P(F|D,Q) \approx 1$.

We use BM25 to search and measure the similarity between two descriptions and develop a rule to judge whether two quantities are the same considering the number of their significant digits and the rounding problem. The overview of the process is shown in Algorithm~\ref{algo:weak_supervision}, where \texttt{SameValue} is the rule-base value-coincidence judging function, and $k$ is the number of results from BM25.

\subsection{Ranking Model}
% \textbf{Probabilistic-based method.} BM25~\cite{robertson2009probabilistic} is one of the most famous ranking functions which utilizes the document and word attributes like term frequency and  document frequency to compute the relevance of a document and a query. 
% It is defined as follows:
% $$score(q, d) = \sum_{t \in q} w_t r(t, d)$$
% where $t$ is a term in query $q$, $w_t$ is the weight of $t$, which is usually defined as Inversed Document Frequency (IDF), $w_t = IDF_t = \log \frac{N-DF_t + 0.5}{DF_t + 0.5}$, where $DF_t$ is the number of documents that contains $t$. A larger IDF means the term is more informative.
% The function $r$ is defined as
% $$r(t, d) = \frac{(k_1 + 1) TF(t, d)}{k_1(1-b+b\frac{L_d}{L_{avg}})+ TF(t, d)}$$
% where $TF(t, d)$ is the term frequency of $t$ in $d$, $L_d$ is the length of $d$ and $L_{avg}$ is the average document length over the whole corpus.
% $b$ determines the scaling by document length, b=1 means using the document length to scaling, and b=0 means ignoring document length. $k_1$ determines the scaling by term frequency, a large value means using raw term frequency, and $k_1=0$ means ignoring term frequency.

% \textbf{Embedding-based method.} One drawback of BM25 is that they rely on textual similarity and may fail to capture the similarity of different expressions of the same thing, like earning and income.
% Pre-trained language models like BERT~\cite{devlin2018bert} have shown their power on several natural language processing tasks to represent the meaning of a sentence or word using a high-dimensional vector.
Based on the paraphrase dataset constructed above, we are able to train a ranking model.
We adopt the Siamese Network~\cite{reimers2019sentence} which encodes a description (or a query) into a dense vector and computes their cosine similarity to measure the similarity between descriptions. Given the vectors of a query and a description $h_x, h_{d_i}$, their similarity is defined as:
% We encode a description into a vector using BERT as follows. Given a description $d=(w_1, ..., w_n)$, we add a special token \texttt{[CLS]} before the description $d'=(\texttt{[CLS]}, w_1, ..., w_n)$, and then feed them into BERT to get embeddings $$(h_{\texttt{[CLS]}}, h_1, ..., h_n)=BERT(\texttt{[CLS]}, w_1, ..., w_n).$$
% There are two ways to give a fixed-size representation $h_d$ of description $d$. The first one is using $h_{\texttt{[CLS]}}$, which is used as the sentence representation in next-sentence-prediction task during pre-training. The second way is taking the average pooling over the token hidden vectors $h_{avg}=\frac{1}{n}\sum_{i=1}^{n} h_i$.
% Similarly, we can encode a query into a vector $h_q$.

% The similarity score of a document and a query is defined as the cosine similarity between their representation vectors:
$$score(x, d_i) = \frac{h_x^T h_{d_i}} { |h_x| \cdot |h_{d_i}| }$$

We use contrastive loss for this task, which pulls in the representations of paraphrases and pushes away the representations of confusing descriptions.
\section{Experiments on Description Parsing}\label{sec:exp-parse}
We collect a large-scale description parsing dataset in the finance area since they are replete with quantities, and the quantity searching demand is urgent in this field.
This section focuses on introducing how to collect the dataset and the challenge of this problem. So, we adopt a de facto standard sequential labeling model based on BERT and do not compare it with other methods.

\subsection{Dataset Collection}

We collect 1,612 documents in published Chinese, including annual reports, IPO prospectuses, bond prospectuses, and industry research reports.
They cover 17 out of 20 categories of industry defined in \emph{Industrial classification for national economic activities}, such as construction, manufacturing, mining, finance, etc.
Since a document such as an annual report has hundreds of pages, it contains information not relevant to quantity retrieval, for example, the table of content, information about the manage teams, etc. So, we focus on chapters mostly composed of factual quantities, such as ``industry status and prospect''.
We implement a rule-based quantity extraction method to extract quantities from these documents. We only retain sentences that contain at least one quantity.
Finally, 42,130 sentences are collected.

We developed a labeling system and trained six in-house annotators to label each quantity's description in these sentences manually. Each sentence is labeled by at least two annotators, and the sentences with conflict results are checked by a third reviewer.
Finally, 112,117 quantities are annotated.

\begin{table*}[t]
\small 
\centering
\caption{Results on Description Parsing.}
\label{tab:parse_result}
\begin{tabular}{lcccccccc}
\toprule
      & \multicolumn{4}{c}{Strict Mode}    & \multicolumn{4}{c}{Partial Mode}   \\
  & \multicolumn{3}{c}{Segment Level} & Quantity Level & \multicolumn{3}{c}{Segment Level} & Quantity Level \\ \cmidrule(lr){2-5} \cmidrule(lr){6-9}
         & Prec. & Rec.  & F1    & Accuracy   & Prec. & Rec.  & F1    & Accuracy   \\
Model-2K\   & 86.81 & 86.06 & 86.44 & 64.21 & 92.26 & 91.43 & 91.83 & 69.70 \\
Model-5K\  & 87.02 & 88.13 & 87.57 & 66.27 & 91.84 & 93.01 & 92.42 & 71.47 \\
Model-full\   & 89.17 & 89.77 & 89.47 & 70.05 & 93.14 & 93.78 & 93.46 & 74.61 \\
Human    & 90.74 & 87.65 & 89.16 & 70.94 & 94.53 & 91.31 & 92.89 & 74.97 \\ \bottomrule
\end{tabular}
\vspace{-1em}
\end{table*}

Now, we introduce some detail of the dataset. Among all quantities, 88\% have descriptions, and 12\% are not related to a fact thus and do not annotate descriptions. The annotated result for each quantity is a set of ``segments'' in the sentence, where each segment is a continuous text that describes some aspect of the quantity. A simple description might include two or three segments, including the time, the subject, and the performance indicator, such as 2021, Tesla, number of employees. But there are many quantities with more segments due to more semantic factors or discontinuous factors, as discussed in Section~\ref{subsec:discuss}. The average number of segments in a quantity description is 3.7, and the maximum number is 10, which indicates that the description can be very complicated.

\subsection{Settings}
The encoder layer of the model uses BERT~\cite{devlin2018bert} and a two-layer Bi-LSTM on top of it. We concatenate the hidden vectors of the last two layers of the BERT-base and feed them into Bi-LSTM, whose hidden size is 512. We also add a dropout layer between Bi-LSTM layers with a ratio of 0.1. We use Adam~\cite{2014Adam} as our optimizer with a learning rate of 2e-5. The dataset is split into training, development, and test sets by 7:1.5:1.5 at the sentence level (so that quantities in the same sentence are not scattered in both training and test sets).

Since we are using sequential labeling for description parsing, we adopt the metric from Named Entity Recognition (NER) for evaluation. Each sentence-quantity pair is regarded as a sample containing several ``entities'' (segments in the description).
We report the results from two granularity, namely, segment and quantity levels. 
At the segment level, we report precision, recall, and F1 to measure how the model performs at extracting useful text pieces to describe a quantity. For example, precision is defined as
$$P=\frac{\sum_{(s,q) \in D} \sum_{e\in P(s,q)} \mathds{1}(e \in Y(s,q)) }{\sum_{(s,q)} \in D |P(s,q)|}$$
where (s, q) is a sentence-quantity pair in dataset $D$, $P(s,q)$ and $Y(s, q)$ are entities in prediction and ground truth, and $\mathds{1}(\cdot)$ is the indicator function that output 1 if $\cdot$ is true, and 0 otherwise.

At the quantity level, we report accuracy, which is the proportion of quantities that we correctly extract their description:
$$Acc= \frac{1}{N}\sum_{(s,q) \in D} \mathds{1}(P(s,q) = Y(s,q)) .$$

Sometimes a minor shift of the boundary of a segment only have a negligible effect on the result, like adding or removing a preposition. NER tasks also regard predicted entities that overlap with ground truth entities as partially correct. So, we also compute the above scores under a ``\emph{partial}'' mode, where a predicted entity is regarded correct if there exists a ground truth entity, such that the length of their intersection is more than 1/3 of their union.

\subsection{Results}

Table~\ref{tab:parse_result} reports the results. The segment-level F1 score reaches 89.47, which is satisfactory. However, the quantity-level accuracy is lower at 70.05.

This drop reflects the difficulty of description parsing in financial text. A quantity description often contains multiple segments, and an error in any segment counts as a failure at the quantity level. The challenge is also evident in human annotations. Each quantity has two annotations; we treat one as “ground truth” and the other as “prediction” to estimate human performance, shown as “Human” in Table~\ref{tab:parse_result}. Despite training and daily meetings, annotators had varied interpretations—particularly on ellipsis, and inclusion of verbs or prepositions. To ensure consistency, a senior annotator resolved all conflicts. Thus, the training data is more consistent than the raw human agreement score suggests.

To test data efficiency, we trained models on 5K and 2K sentence subsets (Model-5K and Model-2K). Even with 2K sentences, the model achieved 86.44 F1, just 3 points below full-data training. As expected, performance improves with more data.

As shown in the next section, even models with modest quantity-level accuracy can substantially enhance retrieval.

\section{Experiments on Quantity Retrieval}

\subsection{Dataset Collection}
\textbf{Quantity Description Corpus $\mathcal{C}_q$.} Based on the description parsing model in Section~\ref{sec:exp-parse}, we collect a large dataset for the quantity retrieval task. 8,845 documents are collected and filtered as in Section~\ref{sec:exp-parse}, and there are 635,194 sentences left. We apply the description parsing model to the resultant sentences and 2,999,378 quantities are parsed with descriptions.

% #docs in train/test: 6191 2654
% #qdicts in train/test: 702059 1000897
% queries:188108 #queries in train/test 132251 55857
% pairs (59568, 12042951)

\textbf{Paraphrase Dataset and Ranking model.} We construct a paraphrase dataset using our proposed Value-Coincidence weak supervision method and the quantity description corpus.

We split $\mathcal{C}_q$ into training and test dataset by documents. Specifically, we select 70\% documents as training dataset $D_{tr}$, generate queries and paraphrases in $D_{tr}$, 30\% documents as test corpus $D_{te}$, generate queries in $D_{te}$ and find paraphrases in $D_{tr} \cup D_{te}$.
Finally, we generate 188,108 high quality queries, 12,042,951 BM25 results, and find 59,568 paraphrase pairs (many queries not find any paraphrases).

We fine-tune a BERT model~\cite{devlin2018bert} (bert-base-chinese) using S-BERT's implementation~\cite{reimers2019sentence}, using online contrastive loss, with cosine similarity distance and 0.5 margin. 
We down-sample the negative samples so that each epoch samples at most 5 negative results for each query.
The learning rate, batch size, and training epoch are 2e-5, 120, and 20. On the test dataset, its F1 score, precision, and recall are 0.82, 0.85, and 0.78 respectively. Notice that this result is evaluated on  the down-sampled test set.

\subsection{Settings}
% Evaluation Metrics: MAP，RECALL

The following methods are compared for quantity retrieval:
\begin{itemize}
    \item $\mathcal{C}_s$-BM25: BM25 method on sentence corpus $\mathcal{C}_s$, where a record in the corpus is a sentence.
    \item $\mathcal{C}_q$-BM25: BM25 method on quantity description corpus $\mathcal{C}_q$.
    \item $\mathcal{C}_q$-BERT: S-BERT method on $\mathcal{C}_q$ using BERT (bert-base-chinese) with no fine tuning. 
    \item $\mathcal{C}_q$-BERT-P: S-BERT method on $\mathcal{C}_q$ using a  multilingual model fine-tuned on other paraphrase detection datasets (paraphrase-multilingual-mpnet-base-v2).
    \item $\mathcal{C}_q$-BERT-WS: S-BERT method on $\mathcal{C}_q$ trained on our proposed weak supervised dataset (the ranking model in the previous subsection).
\end{itemize}
Mean pooling is used to get a fixed-length vector representation for BERT-based methods.

% We compare our QuaRP framework (\emph{Parsing-based}) with a \emph{Sentence-based} retrieval framework whose corpus is composed of sentences.
% \begin{itemize}
%     \item Our \textbf{parsing-based} framework retrieves by parsed descriptions, and each retrieved element is one quantity corresponding to the description. An element is regarded as correct if the quantity is correct. 
%     \item The \textbf{sentence-based} framework retrieves by sentence text. The corpus of this framework is composed of sentences, and each element in the returned result is a set of quantities in the sentence. An element is regarded as correct if any quantity in the set is correct, which means this framework is evaluated \emph{more loosely}.
% \end{itemize}

% For both frameworks, we test them using two similarity measuring methods. The first one is BM25~\cite{robertson2009probabilistic}, we use the default parameters $k_1=1.5$ and $b=0.75$. The second one is an embedding-based method using Sentence-BERT~\cite{reimers2019sentence} and uses its pre-trained model \texttt{paraphrase-multilingual-MiniLM-L12-v2} which maps a text snippet to a 384-dimensional dense vector.

\subsection{Evaluation Method}
\textbf{Automatically generated Test Set.} We use the descriptions in the test dataset of the ranking model for evaluation.
We remove them from the $\mathcal{C}_q$ corpus to avoid trivial results that return the query itself.
For a query $(d_i, q_i)$ and a result $(d_j, q_j)$ in the result list, we set its label as relevant if $q_j$ is the same as $q_j$.
The $\mathcal{C}_s$-BM25 method retrieves at the sentence level, we traverse all the quantities in the sentence, and if any quantity has the same value as $q_i$, we set the sentence as relevant. So, $\mathcal{C}_s$-BM25 has an advantage over $\mathcal{C}_q$-based methods on the evaluation method.
% Since each description in the dataset is coupled with a quantity, the quantity  is used to judge the result.
% Specifically, given a retrieved list of (description, quantity) pairs, if the quantity corresponding to the description is similar to the quantity of the query, we regard the description as relevant, otherwise irrelevant. 
As we discussed before, the chance that two irrelevant descriptions have the same quantity value is small, the relevance judge result is acceptable. And because the chance is random, it is fair for all methods compared.
We pool the results of all methods before evaluation to make a fair comparison. That is, we merge the relevant results of all methods as the relevant result set for each query.
We only retain queries that have at least one relevant result after pooling for evaluation.

\textbf{Manually Annotated Test Set.}
We randomly select 291 queries and their pooling results (all relevant results and part of irrelevant results) and manually annotate their relevance for evaluation.

Then, we introduce the evaluation metrics. As mentioned above, for a query, each element in the retrieved list has a binary relevant label. We use $R(q)$ to denote the relevance result corresponding to query $q$, where $R_i(q)$ is 1 if the i-th element is in fact relevant and 0 otherwise, $R_{1:n}(q)=\{R_1(q), ..., R_n(q)\}$. Denoting all queries in the test set as $Q$.
What we are concerned about most is whether the expected quantity exists in the list, especially at the top of the list. So, we define $Exist@n$ as:
\begin{equation*}
    Exist@n = \frac{\sum_{q \in Q} \mathds{1}( 1 \in R_{1:n}(q) )}{|Q|}
\end{equation*}

While Exist@n measures the existence of the expected quantity in the result, we also want to measure whether relevant results are ranked high. The MAP@n is used to measure this, defined as:
\begin{equation*}
\begin{aligned}
    MAP@n &= \frac{\sum_{q \in Q} AP@n(q)}{|Q|}  \\
    AP@n(q) &= \frac{\sum_{i=1}^{n}P@i(q) \times R_i(q)}{\sum_{k=1}^{n} R_k(q)} \\
    P@n(q) &= \frac{\sum_{i=1}^{n} R_i(q)}{n}. \\
\end{aligned}
\end{equation*}
Another similar measure is nDCG, defined as:
\begin{equation*}
\begin{aligned}
    nDCG@n &= \frac{\sum_{q \in Q} nDCG@n(q)}{|Q|}  \\
nDCG@n(q) &= DCG@n(q) / IDCG@n(q) \\
DCG@n(q) &= \sum_{i=1}^{n} \frac{R_i(q)}{\log_2(i+1)} \\
IDCG@n(q) &= \sum_{i=1}^{|REL@n|} \frac{1}{\log_2(i+1)} \\
\end{aligned}
\end{equation*}
where $|REL@n|$ is the number of relevant results in the corpus up to position n.

% Finally, since a query is usually related to exactly one quantity, we propose a voting method to return one quantity for each query. Specifically, we group quantities so that quantities in a group have the same value and return the group with the largest summation of similarity scores. 
% Thus, we use \emph{Accuracy} to denote the proportion of queries that we can return a correct answer.

\subsection{Results}

% Please add the following required packages to your document preamble:
% \usepackage{booktabs}
\vspace{-0.5em}
\begin{table*}[]
\small
\centering
\caption{Results on Quantity Retrieval}
\label{tab:QR_result}
\begin{tabular}{@{}lcccc|cccc@{}}
\toprule
          & \multicolumn{4}{c}{Automatically Generated Test Set} & \multicolumn{4}{c}{Manually Annotated Test Set} \\ \midrule
          & Exist@1    & Exist@10   & MAP@10   & nDCG@10   & Exist@1    & Exist@10    & MAP@10   & nDCG@10   \\
$\mathcal{C}_s$-BM25 & 0.3098     & 0.5861     & 0.2858   & 0.3521    & 0.2203     & 0.4153      & 0.2071   & 0.2380    \\
$\mathcal{C}_q$-BM25  & 0.5802     & 0.7787     & 0.5064   & 0.5883    & 0.4322     & 0.5847      & 0.3526   & 0.4221    \\
$\mathcal{C}_q$-BERT & 0.5171     & 0.7021     & 0.4266   & 0.5030    & 0.4407     & 0.5508      & 0.3164   & 0.3890    \\
$\mathcal{C}_q$-BERT-P  & 0.5458     & 0.7437     & 0.4620   & 0.5432    & 0.4407     & 0.6102      & 0.3607   & 0.4272    \\
$\mathcal{C}_q$-BERT-WS  & 0.6466     & 0.7839     & 0.5862   & 0.6575    & 0.5000     & 0.6017      & 0.4436   & 0.5004    \\ \bottomrule
\end{tabular}
\vspace{-1em}
\end{table*}

\begin{figure}[t]
\centering
\includegraphics[width=.95\textwidth]{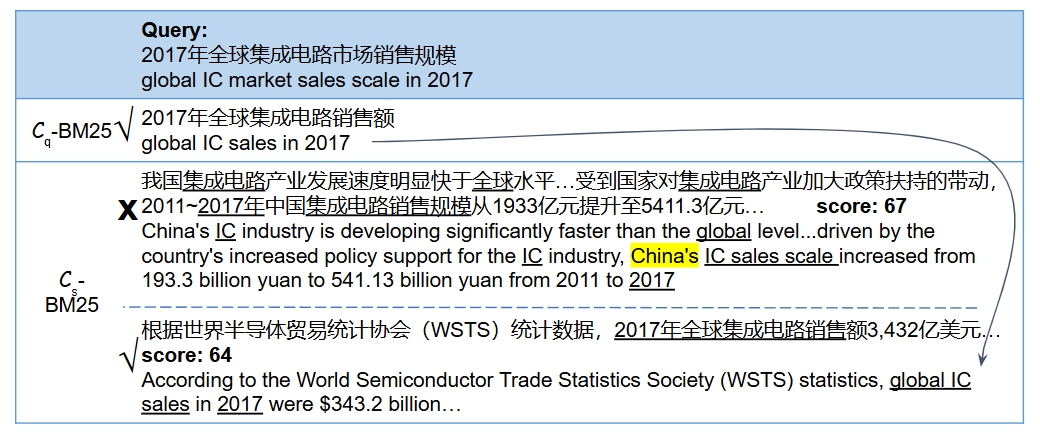}
\caption{A case on which $\mathcal{C}_q$-BM25 return the correct result at top but $\mathcal{C}_s$-BM25 ranks it after a wrong result.}
\label{fig:bm25q-vs-bm25s}
\vspace{-1em}
\end{figure}

The results are shown in Table~\ref{tab:QR_result}. We first discuss the results on the weakly supervised test set and then the results on the manually annotated test set.

\textbf{The Effect of Description Parsing.}
We compare the results of $\mathcal{C}_s$-BM25 and $\mathcal{C}_q$-BM25 to illustrate the improvement brought by description parsing.
% The sentence-based method performs poorly, which returns the correct quantity at the top position on 30\% of cases.
Our proposed parsing-based $\mathcal{C}_q$-BM25 method is significantly better than $\mathcal{C}_s$-BM25 on all metrics. It returns the correct quantity at the top position (Exist@1) on 58\% of cases, which achieves an over 90\% relative improvement compared with $\mathcal{C}_s$-BM25. 
This indicates that retrieving a quantity using its description requires an understanding of its semantic meaning.

A case is shown in Figure~\ref{fig:bm25q-vs-bm25s} to illustrate this. $\mathcal{C}_q$-BM25 returns the correct result on the top. $\mathcal{C}_s$-BM25 returns a wrong result on the top, and the correct one is ranked in the second position. This is because the sentence in the first result includes all the query terms, and important terms like ``IC'' are repeated multiple times. 
But the quantities in this sentence are about the Chinese market (highlighted in the figure) although the term ``global'' appears in the sentence. 
For $\mathcal{C}_q$-BM25, the descriptions of these quantities do not contain the term ``global'', thus are ranked lower.

\begin{figure}[t]
\centering
\includegraphics[width=\textwidth]{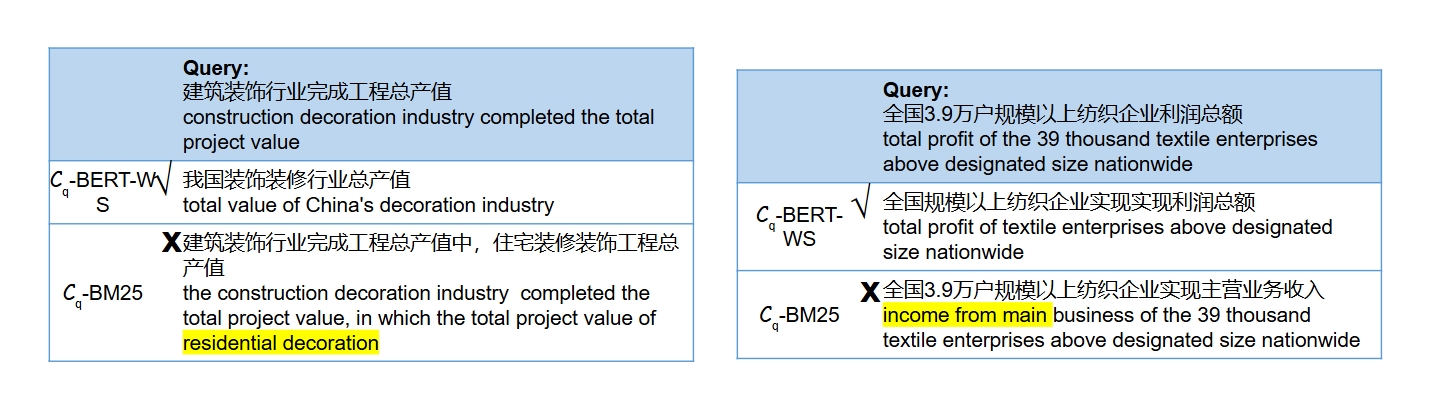}
\caption{Two cases on which $\mathcal{C}_q$-BERT-WS performs better than $\mathcal{C}_q$-BM25 when comparing their top results.}
\label{fig:bm25q-vs-ws}
\vspace{-1em}
\end{figure}

% For example, in ``BM25'' lines, the parsing-based framework outperforms sentence-based under all evaluation metrics with around 50\% improvements. And the improvement is consistent under different metrics. The poor performance  of the sentence-based framework indicates that  the quantity retrieval task is a difficult task, and existing methods have limitations on this task. Our proposed parsing-based framework can help to overcome the limitations.
% Then we discuss the results in detail. 

\textbf{The Effect of Weak Supervision.} We compare the results of $\mathcal{C}_q$-BM25 and $\mathcal{C}_q$-BERT-WS. It shows that training a neural network model using the proposed value-coincidence weak supervision method can further bring a 6\% absolute improvement on Exist@1.  This means the ranking model can learn paraphrase information from the weak supervision dataset.

Two cases are shown in Figure~\ref{fig:bm25q-vs-ws} to illustrate this. Each case shows its query and the top-ranked result for each method.
In the first case, $\mathcal{C}_q$-BERT-WS is able to infer that ``construction decoration industry'' is the same as ``decoration industry'', ``total project value'' is the same ``total value'' in this industry, and  this is usually a nationwide industry statistic if not specified, thus adding ``China'' does not influence its meaning (since this is a Chinese annual report corpus).
The top result of $\mathcal{C}_q$-BM25 includes all the terms in the query, so it gets a high score. But these terms are followed by a more specific phrase, indicating it only measures a part of the decoration industry: the residential decoration sub-industry (there are other sub-industries like public building decoration).
In the second case, the query includes an unnecessary term ``39 thousand'' which is the number of textile enterprises above the designated size. $\mathcal{C}_q$-BERT-WS is able to learn that this term is not important.
But since this term appears rarely in the dataset, but financial indicators like total profit and income from main business appear frequently in this annual report corpus, $\mathcal{C}_q$-BM25 give more weights on the term ``39 thousand'' thus return the wrong description as the top result.
These cases illustrate the effect of weak supervision on understanding what is a  paraphrase.

We also compare $\mathcal{C}_q$-BERT-WS with $\mathcal{C}_q$-BERT and $\mathcal{C}_q$-BERT-P. The latter two perform worse than $\mathcal{C}_q$-BERT-WS and even worse than $\mathcal{C}_q$-BM25. This means \textbf{the improvement comes from the weakly supervised dataset}, rather than the knowledge in the pre-trained language model. We think $\mathcal{C}_q$-BERT-P performs worse than $\mathcal{C}_q$-BERT because the fine-tuning datasets have different distributions with this application.

Notice that, the improvement of $\mathcal{C}_q$-BERT-WS on Exist@10 to $\mathcal{C}_q$-BM25 is marginal compared with the improvement on other metrics. It indicates that $\mathcal{C}_q$-BM25 can  recall most cases if we go down the result list, although they are ranked after some confusing descriptions. And the fact that both of them perform better than $\mathcal{C}_s$-BM25 reflects the effectiveness of description parsing.

\begin{table}[]
\caption{The ``win-matrix'' measuring the percentage of winning cases between model pairs (on nDCG)}
\label{tab:win-matrix}
\centering
\begin{tabular}{@{}lccccc@{}}
\toprule
                        & \begin{tabular}[c]{@{}c@{}}$\mathcal{C}_s$-\\ BM25\end{tabular} & \begin{tabular}[c]{@{}c@{}}$\mathcal{C}_q$-\\ BM25\end{tabular} & \begin{tabular}[c]{@{}c@{}}$\mathcal{C}_q$-\\ BERT\end{tabular} & \begin{tabular}[c]{@{}c@{}}$\mathcal{C}_q$-\\ BERT-P\end{tabular} & \begin{tabular}[c]{@{}c@{}}$\mathcal{C}_q$-\\ BERT-WS\end{tabular} \\ \midrule
$\mathcal{C}_s$-BM25    &                                                                 & 10\%                                                            & 18\%                                                            & 12\%                                                              & 8\%                                                                \\
$\mathcal{C}_q$-BM25    & 44\%                                                            &                                                                 & 20\%                                                            & 15\%                                                              & 5\%                                                                \\
$\mathcal{C}_q$-BERT    & 40\%                                                            & 14\%                                                            &                                                                 & 12\%                                                              & 6\%                                                                \\
$\mathcal{C}_q$-BERT-P  & 44\%                                                            & 19\%                                                            & 24\%                                                            &                                                                   & 6\%                                                                \\
$\mathcal{C}_q$-BERT-WS & 45\%                                                            & 31\%                                                            & 32\%                                                            & 29\%                                                              &                                                                    \\ \bottomrule
\end{tabular}
\vspace{-1em}
\end{table}

\textbf{Results on Manually Annotated Test Set.} 
The results on the manually annotated test set are similar to the automatically generated test set: $\mathcal{C}_q$-BERT-WS performs better than $\mathcal{C}_q$-BM25, and $\mathcal{C}_q$-BM25 performs better than $\mathcal{C}_s$-BM25. But the scores of all these methods are dropped. This is because the manually annotated test set is more strict on deciding what is a relevant result. We also compute a ``win-matrix'' $W$ on this dataset, where $W_{i}{j}$ is the percentage of the cases that the i-th method performs better than the j-th method on nDCG. Table~\ref{tab:win-matrix} shows the result. We can see that $\mathcal{C}_q$-BERT-WS performs better on most samples. We manually check the samples on which $\mathcal{C}_q$-BERT-WS performs worse than $\mathcal{C}_s$-BM25 and find most of them are due to description parsing errors.

\section{Related Work}
We categorize related work into two areas: quantity-based information retrieval and information extraction.

\textbf{Quantity retrieval.}  
Banerjee et al.~\cite{banerjee2009learning} introduced quantity consensus queries, where each query returns a quantity interval (e.g., “driving time from Paris to Nice” → 15–20 minutes). Their method ranks intervals using surrounding snippets as feature vectors, akin to our sentence-based baseline, but lacks deep semantic understanding.

Sarawagi and Chakrabarti~\cite{sarawagi2014open} addressed quantity retrieval in web tables, assuming semantics are conveyed via row/column headers. They also proposed a consensus-based reranking model. In contrast, our work handles unstructured text, which poses greater semantic challenges and allows broader application.

Ho et al.~\cite{ho2019qsearch,ho2020entities} developed Qsearch for entity retrieval under quantity constraints (e.g., “hybrid cars under 35,000 Euros”), while Li et al.~\cite{li2021anasearch} proposed AnaSearch for answering analytical questions over COVID-19 data (e.g., “total deaths in Italy by April 21st”). Both use syntactic parsing for structure, whereas we apply deep learning for richer representation learning.

\textbf{Quantity understanding.}  
Lamm et al.~\cite{lamm2018qsrl} proposed a semantic role labeling schema for financial quantities, identifying 11 distinct roles but lacking data or experiments. Saha et al.~\cite{saha2017bootstrapping} introduced BONIE, a bootstrapped Open IE system for extracting tuples with numeric arguments, relying on syntactic patterns that struggle with complex expressions.

Thawani et al.~\cite{thawani2021representing} surveyed numeric representation in NLP across tasks like arithmetic~\cite{geva2020injecting}, numeration~\cite{naik2019exploring}, comparison~\cite{wallace2019nlp}, and math word problems~\cite{hendrycks2021measuring,cao2021bottom,xie2019goal}, discussing methods for encoding and decoding numbers in text.

In summary, while prior work has addressed quantities in both structured and unstructured data, we are the first to formally study quantity retrieval from text via natural language queries.

\section{Discussion}\label{subsec:discuss}

\textbf{Limitations and future work.}  
Our framework for quantity retrieval is preliminary, and each module has room for improvement. We highlight two key directions:

First, quantity understanding could benefit from the advances of large language models (LLMs), and replacing our BERT-based parsing method with LLMs~\cite{navigli-etal-2024-nounatlas,10666757} may improve the overall performance of our framework. Moreover, descriptive factors may span sentences or even paragraphs, requiring document-level information extraction~\cite{zheng2019doc2edag,xu2021document}.

Second, paraphrase detection is vital for retrieval. We proposed a basic weakly supervised method based on value overlap. Future work could explore abbreviation resolution and learning from noisy weak labels\cite{qiu-zhang-2024-label}.

% point out some scientific issues.
% The description parsing is a long term task, including argument scatter problem
% discontinuous NER
% weak supervision considering inter and intra doc
% the abbr problem

\section{Conclusion}

In this paper, we propose the task of quantity retrieval from text. This paper argues that this task requires a semantic understanding of quantities. By analyzing the challenge, we propose a description parsing-based framework with weak supervision, which significantly outperforms the sentence-based framework. We also summarize some of the future research directions on this task.

\bibliographystyle{splncs04}
\bibliography{mypaper}

\begin{thebibliography}{10}
\providecommand{\url}[1]{\texttt{#1}}
\providecommand{\urlprefix}{URL }
\providecommand{\doi}[1]{https://doi.org/#1}

\bibitem{banerjee2009learning}
Banerjee, S., et~al.: Learning to rank for quantity consensus queries. In: SIGIR (2009)

\bibitem{10666757}
Cai, W., Li, Y., Chen, Y., Lin, J., Huang, Z., Gao, P., Gadekallu, T.R., Wang, W., Gao, Y.: Enhancing weakly supervised semantic segmentation with multi-label contrastive learning and llm features guidance. IEEE Journal of Biomedical and Health Informatics  (2024)

\bibitem{cao2021towards}
Cao, R., et~al.: Towards document panoptic segmentation with pinpoint accuracy: Method and evaluation. In: ICDAR (2021)

\bibitem{cao2021bottom}
Cao, Y., et~al.: A bottom-up dag structure extraction model for math word problems. In: AAAI (2021)

\bibitem{devlin2018bert}
Devlin, J., et~al.: Bert: Pre-training of deep bidirectional transformers for language understanding. arXiv preprint arXiv:1810.04805  (2018)

\bibitem{geva2020injecting}
Geva, M., et~al.: Injecting numerical reasoning skills into language models. In: ACL (2020)

\bibitem{hendrycks2021measuring}
Hendrycks, D., et~al.: Measuring mathematical problem solving with the math dataset. arXiv:2103.03874  (2021)

\bibitem{ho2019qsearch}
Ho, V.T., et~al.: Qsearch: Answering quantity queries from text. In: ISWC (2019)

\bibitem{ho2020entities}
Ho, V.T., et~al.: Entities with quantities: Extraction, search, and ranking. In: WSDM (2020)

\bibitem{kamathsurvey}
Kamath, A., Das, R.: A survey on semantic parsing. In: AKBC (2019)

\bibitem{2014Adam}
Kingma, D.P., et~al.: Adam: A method for stochastic optimization. arXiv preprint arXiv:1412.6980  (2014)

\bibitem{lamm2018qsrl}
Lamm, M., et~al.: Qsrl: A semantic role-labeling schema for quantitative facts. In: FNP (2018)

\bibitem{li2020cross}
Li, K., et~al.: Cross-domain document object detection: Benchmark suite and method. In: CVPR (2020)

\bibitem{li2021anasearch}
Li, T., et~al.: Anasearch: Extract, retrieve and visualize structured results from unstructured text for analytical queries. In: WSDM (2021)

\bibitem{naik2019exploring}
Naik, A., et~al.: Exploring numeracy in word embeddings. In: ACL (2019)

\bibitem{navigli-etal-2024-nounatlas}
Navigli, R., Lo~Pinto, M., Silvestri, P., Rotondi, D., Ciciliano, S., Scir{\`e}, A.: {N}oun{A}tlas: Filling the gap in nominal semantic role labeling. In: ACL (2024)

\bibitem{qiu-zhang-2024-label}
Qiu, X.Y., Zhang, J.: Label confidence weighted learning for target-level sentence simplification. In: Al-Onaizan, Y., Bansal, M., Chen, Y.N. (eds.) EMNLP (2024)

\bibitem{reimers2019sentence}
Reimers, N., et~al.: Sentence-bert: Sentence embeddings using siamese bert-networks. In: EMNLP (2019)

\bibitem{saha2017bootstrapping}
Saha, S., et~al.: Bootstrapping for numerical open ie. In: ACL (2017)

\bibitem{sarawagi2014open}
Sarawagi, S., et~al.: Open-domain quantity queries on web tables: annotation, response, and consensus models. In: KDD (2014)

\bibitem{thawani2021representing}
Thawani, A., et~al.: Representing numbers in nlp: a survey and a vision. In: NAACL (2021)

\bibitem{wallace2019nlp}
Wallace, E., et~al.: Do nlp models know numbers? probing numeracy in embeddings. In: EMNLP-IJCNLP (2019)

\bibitem{xie2019goal}
Xie, Z., et~al.: A goal-driven tree-structured neural model for math word problems. In: IJCAI (2019)

\bibitem{xu2021document}
Xu, W., et~al.: Document-level relation extraction with reconstruction. In: AAAI. vol.~35 (2021)

\bibitem{yang2018design}
Yang, J., et~al.: Design challenges and misconceptions in neural sequence labeling. In: COLING (2018)

\bibitem{zheng2019doc2edag}
Zheng, S., et~al.: Doc2edag: An end-to-end document-level framework for chinese financial event extraction. In: EMNLP-IJCNLP (2019)

\end{thebibliography}

\end{document}